\documentclass[aip,jcp,twocolumn,superscriptaddress,groupedaddress]{revtex4}  
\usepackage{amsfonts}
\usepackage{amsmath}
\usepackage{amssymb}
\usepackage{version}
\usepackage{graphicx}%
\includeversion{New_connection}
\excludeversion{Old_connection}

\begin{document}
\title{The G-JF Thermostat for Accurate Configurational Sampling in Soft-Matter Simulations}
\author{Evyatar Arad}
\affiliation{Department of Biomedical Engineering, Ben Gurion University of the Negev, Be'er Sheva, 84105 Israel}
\author{Oded Farago}
\affiliation{Department of Biomedical Engineering, Ben Gurion University of the Negev, Be'er Sheva, 84105 Israel}
\affiliation{Ilse Katz Institute for Nanoscale Science and Technology, Ben Gurion University of the Negev, Be'er Sheva, 84105 Israel}
\author{Niels Gr{\o}nbech-Jensen}
\affiliation{Department of Mechanical and Aerospace Engineering, University of California, Davis, CA 95616}
\affiliation{Department of Mathematics, University of California, Davis, CA 95616}
\begin{abstract}
We implement the statistically sound G-JF thermostat for Langevin Dynamics simulations into the ESPREesSo molecular package for large-scale simulations of soft matter systems. The implemented integration method is tested against the integrator currently used by the molecular package in simulations of a fluid bilayer membrane. While the latter exhibits deviations in the sampling statistics that increase with the integration time step $dt$, the former reproduces near-correct configurational statistics for all $dt$ within the stability range of the simulations. We conclude that, with very modest revisions to existing codes, one can significantly improve the performance of statistical sampling using Langevin thermostats.
\end{abstract}
\maketitle

\section{Introduction}
\label{sec:intro}
Discrete-time Molecular Dynamics (MD) is one of the most common methods for simulating molecular systems \cite{Rapaport}. MD simulations are performed by numerically integrating NewtonÕs equations of motion to advance the coordinates of the particles in discrete time. The most frequently used numerical integrator for performing MD is based on the St{\o}rmer-Verlet algorithm \cite{Verlet}, which, for a closed system, can be written
\begin{eqnarray}
r^{n+1} & = & r^n + dt\,v^n+\frac{dt^2}{2m}f^n \label{eq:Verlet_r} \\
v^{n+1} & = & v^n + \frac{dt}{2m}(f^n+f^{n+1}) \, , \label{eq:Verlet_v}
\end{eqnarray}
where $r^n$, $v^n$, and $f^n=f(r^n)$ denote the position, velocity, and force of a particle, respectively, at time $t_n$. The St{\o}rmer-Verlet algorithm is a second order integrator in $dt$. It is considered favorable over other, higher order in $dt$, integrators due to its simplicity, computational efficiency, and global conservation properties. For a closed system, these properties ensure optimal stability, time reversibility and, e.g., effective energy conservation over long time integrations \cite{Frenkel}.
Despite these attractive features, a little-appreciated fact is that the parameter we assign to represent
velocity (or momentum) in discrete-time dynamics is not exactly the conjugated variable to the simulated position.
The conjugated relationship between the position and velocity coordinates is recovered only in continuous time (see Appendix A in Ref.~\cite{GJF3} and references therein). The consequences of this fundamental artifact are significant as one must accept that kinetic and configurational measures cannot be obtained correctly from the same simulation unless a simulation is conducted with a very small time step.

The St{\o}rmer-Verlet algorithm addresses dynamics in closed (microcanonial) systems, characterized by conservation of the total energy. However, the microcanonial ensemble is less relevant for most MD applications than its canonical counterpart, where the temperature, rather than the energy, is constant. This is especially true for the relatively small systems that often are simulated as a proxy for thermodynamically large ensembles. A number of methods for constraining the temperature of a simulated system (ÒthermostatsÓ) exists; two representative ones are the deterministic (e.g., Nos{\'e}-Hoover \cite{Nose,Hoover}) and stochastic (Langevin) thermostats \cite{Paquet}. Here we focus on integration methods for Langevin Dynamics (LD). In LD, two terms are added to Newton's equations of motion: (i) friction proportional and opposite  to the velocity, and (ii) an accompanying delta-function correlated (ÒwhiteÓ) thermal noise. Langevin's equation is, thus, given by \cite{Langevin}:
\begin{eqnarray}
\dot{r} & = & v \label{eq:L_r} \\
m\dot{v} & = & f(r,t)-\alpha v + \beta(t) \, , \label{eq:L_v}
\end{eqnarray}
where $f(r,t)$ is the deterministic force acting on the particle, $\alpha>0$  is a constant friction coefficient, and   $\beta(t)$ denotes the thermal noise. In order to satisfy Einstein's fluctuation-dissipation theorem, it can be assumed that the noise is Gaussian-distributed, with the following statistical properties \cite{Parisi}:
\begin{eqnarray}
\langle\beta(t)\rangle & = & 0 \label{eq:noise_cont_ave} \\
\langle\beta(t)\beta(t^\prime)\rangle & = & 2\alpha k_BT\delta(t-t^\prime) \, , \label{eq:noise_cont_std}
\end{eqnarray}
where $k_B$ is Boltzmann's constant and $T$ is the thermodynamic temperature.

Developing an accurate numerical integrator for Langevin's equation is not trivial due to the non-analytic nature of the thermal noise, and the fact that the friction force is velocity-dependent. If the friction and noise terms are treated on equal footing with $f(r,t)$, one obtains the frequently-used BBK (Br{\"u}nger, Brooks, Karplus) integrator, which is simple, yet known to be inaccurate when employed with moderate to large integration time step $dt$ \cite{BBK}. Specifically, the BBK, as well as
most
other existing integrators
(including Nos{\'e}-Hoover),
tend to exhibit increasing artificial changes in the configurational sampling statistics as the time step is enlarged.
This is rooted, in part, in the above-mentioned discrete-time artifact that momentum and position are not strictly mutually conjugated variables for $dt>0$.
Recently, a new and improved thermostat (a temporal discrete-time propagator of the Langevin equation) was introduced by Gr{\o}nbech-Jensen and Farago (G-JF) \cite{GJF1}, which reads
\begin{eqnarray}
r^{n+1} & = & r^n + b [dt\, v^n+\frac{dt^2}{2m}f^n+\frac{dt}{2m}\beta^{n+1}] \label{eq:gjf_r} \\
v^{n+1} & = & a\, v^n+\frac{dt}{2m}(af^n+f^{n+1})+\frac{b}{m}\beta^{n+1} \, , \label{eq:gjf_v}
\end{eqnarray}
where
\begin{eqnarray}
a & = & \frac{\displaystyle{1-\frac{\alpha dt}{2m}}}{\displaystyle{1+\frac{\alpha dt}{2m}}} \label{eq:a} \\
b & = & \frac{\displaystyle{1}}{\displaystyle{1+\frac{\alpha dt}{2m}}} \, . \label{eq:a}
\end{eqnarray}
The discrete-time noise is
\begin{eqnarray}
\beta^{n+1} & = & \int_{t_n}^{t_{n+1}}\beta(t^\prime)\,dt^\prime \, , \label{eq:discrete_beta}
\end{eqnarray}
which results in an uncorrelated Gaussian random number with zero mean and a variance given by the temperature and friction coefficient:
\begin{eqnarray}
\langle\beta^n\rangle & = & 0 \label{eq:noise_dis_ave} \\
\langle\beta^n\beta^l\rangle & = & 2\alpha k_BT dt \delta_{n,l} \, . \label{eq:noise_dis_std}
\end{eqnarray}
Notice that the limiting case, $\alpha=0$, of the G-JF method outlined in Eqs.~(\ref{eq:gjf_r})-(\ref{eq:noise_dis_std}) reduces the method to the standard St{\o}rmer-Verlet algorithm of Eqs.~(\ref{eq:Verlet_r}) and (\ref{eq:Verlet_v}).

The core of the G-JF method is that the fluctuation-dissipation relationship is intact in discrete-time with respect to the balance between the energy lost by friction over the actual distance traveled and the accumulated noise over the time step\cite{GJF1}. This implies that the resulting discrete-time {\it trajectory} is thermodynamically sound.  It therefore enables simulations of diffusion and {\it configurational} space without compromising the sampling statistics as the time step is varied throughout the numerical stability range \cite{GJF1,GJF2}.
The objective of this paper is to illuminate the statistical performance of the method for both low-dimensional systems as well as complex, soft-matter systems for which we have implemented the G-JF algorithm in the simulation suite, ESPREesSo, in order to demonstrate the resulting improvements that can be attained by modest revisions to existing molecular dynamics codes.

\section{Application to Simple Oscillators}
In order to appreciate the sampling strength of the G-JF thermostat, we first study a particle moving in  one-dimensional space with potential energy $U(r)$.
We investigate the equilibrium statistics of the system by integrating Eqs.~(\ref{eq:L_r}) and (\ref{eq:L_v}) with
\begin{eqnarray}
f(r) & = & -\frac{\partial U}{\partial r} \, . \label{eq:Potential_general}
\end{eqnarray}
We can regard Eqs.~(\ref{eq:L_r}) and (\ref{eq:L_v}) as normalized, along with all variables, if we assume that $r$ is normalized to
a characteristic displacement $r_0$, $m$ is measured in units of $m_0$, energy $U$ in units of $E_0$, time
$t$ in units of $t_0=r_0\sqrt{m_0/E_0}$, velocity $v$ in units of $v_0=r_0/t_0$, and normalized temperature is given by $\theta=k_BT/E_0$. Our simulation results shown in this paper are for $m=\theta=1$.

In what follows, we consider confining potentials, where an object with coordinate $r$ has a localized equilibrium distribution function $\rho_{eq}\sim\exp[-U(r)/k_BT]$.
We simulate the Langevin dynamics with three discrete-time algorithms: G-JF \cite{GJF1}, BBK \cite{BBK}, and one
by Stoll and Schneider (SS) \cite{SS} (where we have set the algorithm parameter $p=1$). The reason for the two latter choices is that they represent commonly used methods in distributed MD suites. From the simulations we obtain the normalized distribution function $\rho(r)$, which we use to generate the normalized potential of mean force $U_{\rm pmf}(r)=c-\theta\ln\rho(r)$,
where $c$ is a constant. A measure of the quality of the applied algorithm is then the difference $U_{\rm pmf}(r)-U(r)$, with an appropriate choice of the constant $c$. We use a total of $10^{10}$ time steps for each acquired distribution function. From these simulations we also derive the
important normalized configurational temperature $\theta_C=k_BT_C/E_0$ \cite{Hirschfelder,Rickayzen},
\begin{eqnarray}
T_C & = & \frac{E_0}{k_B}\frac{\langle(\partial U/\partial r)^2\rangle}{\langle\partial^2U/\partial r^2\rangle} \, , \label{eq:TC}
\end{eqnarray}
which is a
condensed measure of how well the configurational space is sampled.

\begin{figure}[t]
\begin{flushleft}
\scalebox{0.5}{\centering \includegraphics{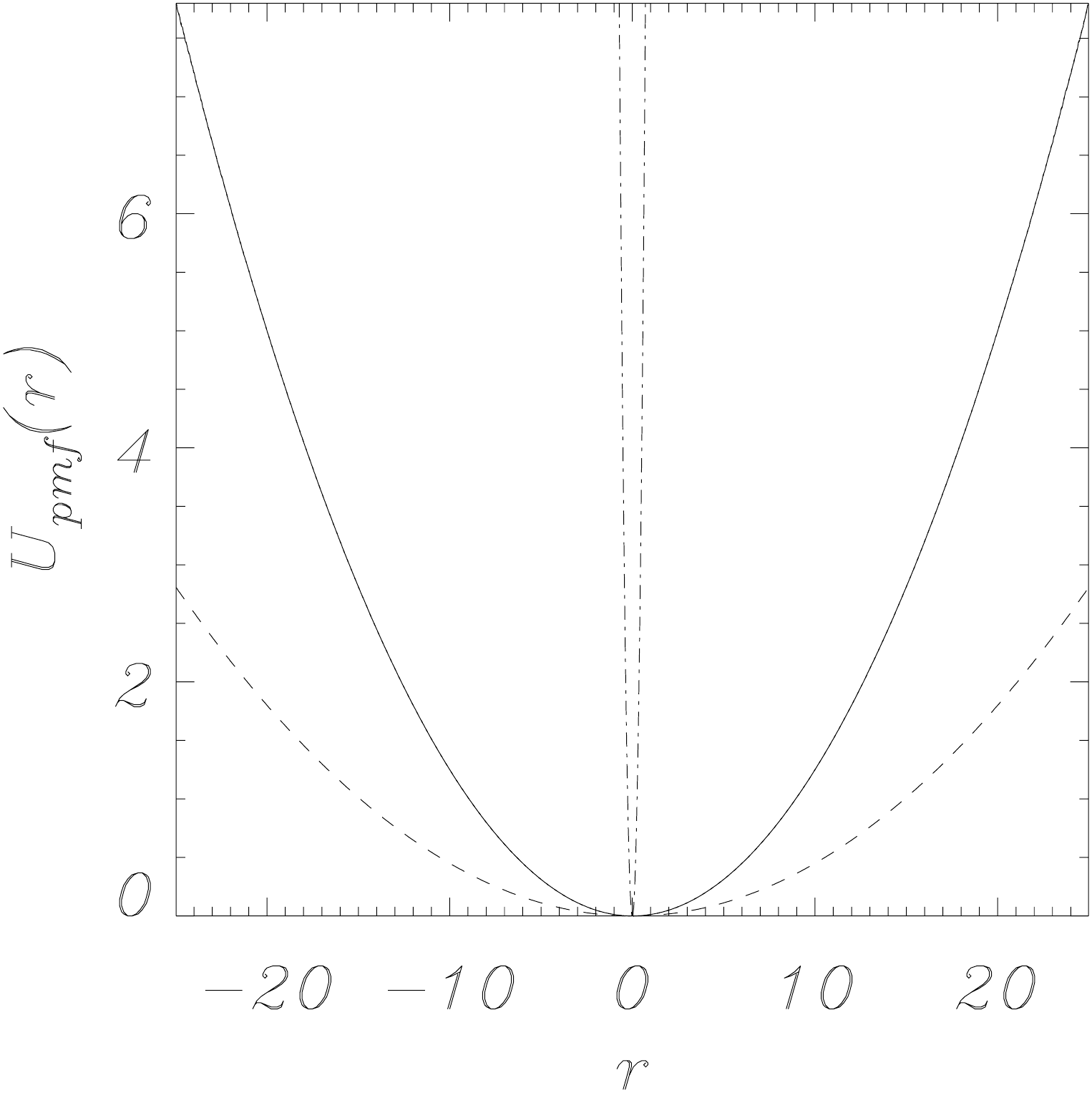}}
\end{flushleft}
\vspace{-0.5 cm}
\caption{Potentials of mean force $U_{\rm pmf}(r)$ from simulated harmonic oscillator using G-JF (solid), BBK (dashed), and SS (dash-dotted) methods for $\alpha=2$ and $dt=0.8dt_{\max}$. True
potential (dotted) is precisely reproduced by G-JF.}
\label{fig:fig_1}
\end{figure}

\begin{figure}[t]
\begin{flushleft}
\scalebox{0.5}{\centering \includegraphics{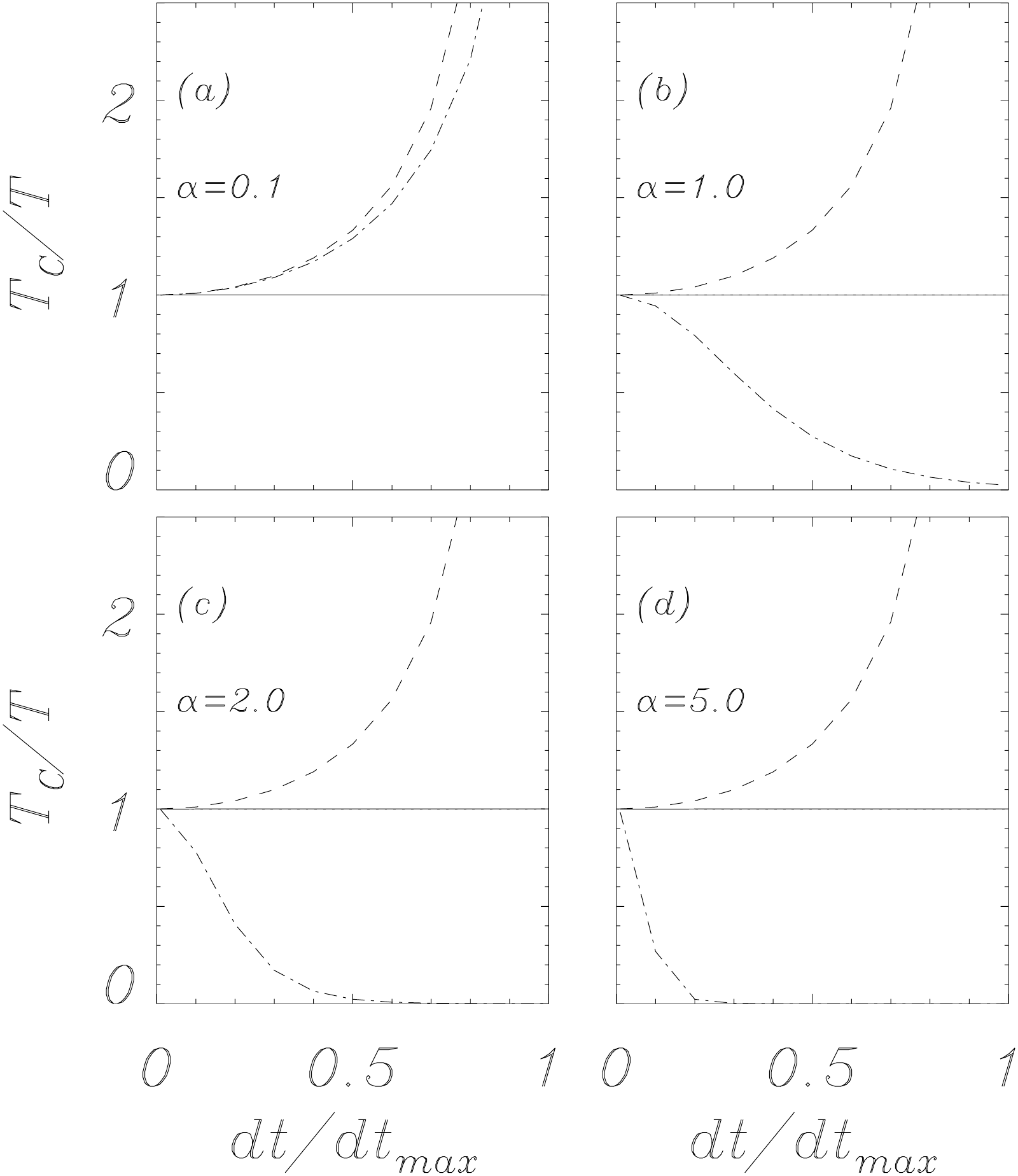}}
\end{flushleft}
\vspace{-0.5 cm}
\caption{Configurational temperature $T_C$ from simulated harmonic oscillator using G-JF (solid), BBK (dashed), and SS (dash-dotted) methods as a function of the applied time step ($dt_{\rm max}$ is the defined stability limit). True
temperature (dotted) $\theta=1$ is precisely reproduced by G-JF.}
\label{fig:fig_2}
\end{figure}

We first consider a harmonic oscillator $U(r)=\frac{1}{2}\kappa r^2$ with $\kappa=1/40$. Since this results in a linear equation of motion, the use of a Gaussian random variable will result in a Gaussian distribution $\rho(r)$ with
zero mean (by symmetry, since $\langle\beta\rangle=0$). It was shown analytically in Ref.~\cite{GJF1} that
the resulting variance of $\rho(r^n)$ is $V(\rho(r))=\frac{1}{2}\theta$, which implies that the G-JF algorithm
reproduces the correct Boltzmann distribution precisely in discrete time for any applied time step $dt<2/\sqrt{\kappa/m}=dt_{\rm max}$ within the stability limit of the extended St{\o}rmer-Verlet methods for Langevin dynamics \cite{ML}. This essential feature is verified by simulations, as shown in Figures 1 and 2, where we display
$U(r)$ along with $U_{\rm pmf}(r)$ (Fig.~1)  and $T_C$, for different values of $\alpha$ and $dt$ (Fig.~2) computed using the three
integration methods mentioned above. We observe the expected perfect agreement between the G-JF results for $U_{\rm pmf}(r^n)$ and $U(r)$ (G-JF results are shown with a solid curve, while $U(r)$, which is
shown dotted, is completely overlapped by the solid curve). In contrast, Fig.~2 shows considerable deviations for both BBK (dashed) and SS (dash-dotted) methods as $dt$ is increased. It is obvious that BBK consistently overestimates the configurational
temperature, which is consistent with the flattening of the effective (pmf) potential that is seen in Fig.~1.
The SS method, however, has a more complex set of errors. For small dissipation, we see that this method
also overestimates $T_C$, while large $\alpha$ generally underestimates the temperature. This is consistent with the hardening of the effective potential observed in Fig.~1 for the SS algorithm.  The results imply that both BBK
and SS methods should be applied with considerable caution, and only with very small time steps compared to the stability limit.

\begin{figure}[t]
\begin{flushleft}
\scalebox{0.5}{\centering \includegraphics{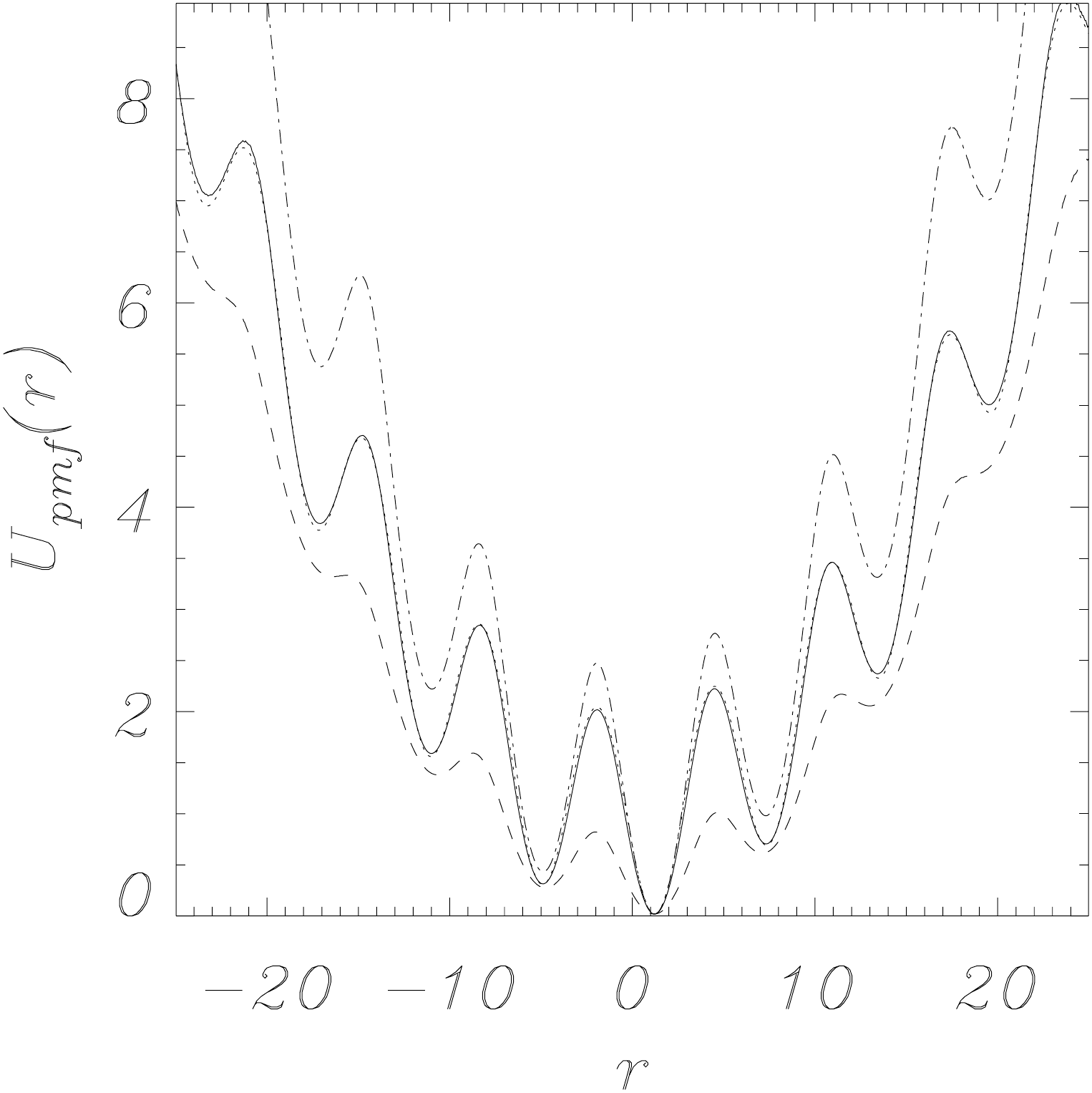}}
\end{flushleft}
\vspace{-0.5 cm}
\caption{Potentials of mean force $U_{\rm pmf}(r)$ from simulated harmonic oscillator using G-JF (solid), BBK (dashed), and SS (dash-dotted) methods for $\alpha=2$ and $dt=0.8dt_{\max}$. True
potential (dotted) is closely reproduced by G-JF.}
\label{fig:fig_3}
\end{figure}

\begin{figure}[t]
\begin{flushleft}
\scalebox{0.5}{\centering \includegraphics{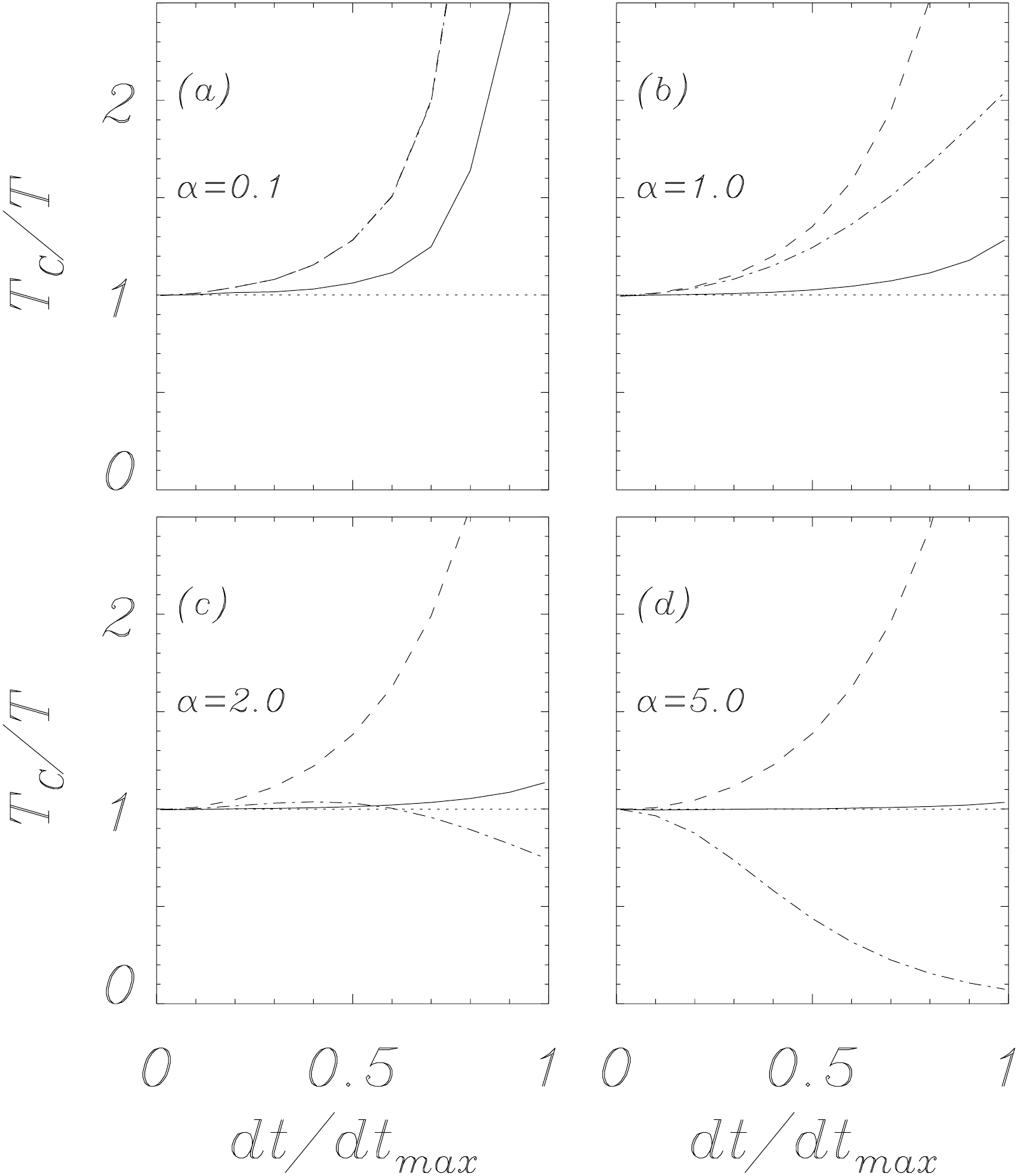}}
\end{flushleft}
\vspace{-0.5 cm}
\caption{Configurational temperature $T_C$ from simulated nonlinear oscillator using G-JF (solid), BBK (dashed), and SS (dash-dotted) methods as a function of the applied time step ($dt_{\rm max}$ is the defined stability limit). True
temperature is shown as a dotted line.}
\label{fig:fig_4}
\end{figure}

Second, we validate the performance of the methods for a highly nonlinear potential $U(r)=\frac{1}{2}\kappa r^2-\cos(r-\xi)$ for $\kappa=1/40$ and $\xi=\frac{3}{4}\pi$ (which is chosen, somewhat arbitrarily, in order to create some asymmetry in the potential). The stability limit of the St{\o}rmer-Verlet methods is given by the maximum curvature of the potential, which in this case is $\tilde\kappa=\kappa+1$. Thus, we define the stability limit for the nonlinear problem to be $dt_{\rm max}=2/\sqrt{\tilde\kappa/m}$. The simulations, which are performed with $\alpha=2$, reconfirm that the intuition from the harmonic oscillator generally translates to the strongly nonlinear case. The
BBK integrator overestimates the configurational temperature by effectively lowering the local energy barriers, and
where the SS method continues to have complex responses to variations in $\alpha$ and $dt$. The G-JF method
is no longer exact when compared to the true (continuous time) expectations, but it is clearly superior compared
to the reference methods. It is important to note that the discrepancies for nonlinear systems arise not from the G-JF method's implementation of dissipation and fluctuations, which are correctly balanced in discrete-time, but is an artifact
of the discrete-time approximations to the behavior of the deterministic force within a single time step. This
is an unavoidable feature common and inherent to all Verlet-type methods.
The observed G-JF trend that the configurational temperature becomes increasingly more accurate for increasing friction coefficient $\alpha$ is due to the fact that the dissipation and fluctuation terms in the Langevin equation (\ref{eq:L_v}) become dominant for large $\alpha$. Figure \ref{fig:fig_4} therefore confirms the desired thermodynamic G-JF properties, since the G-JF method provides the correct configurational dissipation-fluctuation relationship in discrete-time.
Figures 3 and 4 also demonstrate
that what may look to be minor differences in configurational temperature (see Fig.~4c for $dt=0.8dt_{\max}$)
can in fact be masking rather large and significant deviations in the Boltzmann distribution (seen in Fig.~3). This
emphasizes the importance of validating the actual configurational distribution when considering if computer simulations
represent the thermodynamic situation under investigation. Notice, however, that such validation is only possible in low-dimensional systems.
It is important to re-emphasize that kinetic and configurational measures cannot be simultaneously correct, since the velocity parameter in the numerical methods is not exactly the velocity of the simulated trajectory. Thus, while the G-JF method will not provide the expected kinetic temperature, as measured by the average kinetic energy, the reason is that the velocity parameter is, in fact, not consistent with the configurational behavior in discrete time. The interesting complement to this observation is that a simulation method that provides correct kinetic behavior (such as kinetic temperature) cannot also reproduce correct configurational response unless the time step is very small.

\section{Application to Soft-Matter}

The option to employ the G-JF thermostat has been added to the software simulation suite LAMMPS (Large-Scale Atomic/Molecular Massively Parallel Simulator), a popular MD simulator for materials modeling, developed and maintained by Sandia National Laboratories \cite{Plimpton}. The LAMMPS suite has the SS method as its other thermostat option (see MD comparison between G-JF and SS methods in section II above and in Ref.~\cite{GJF2}). Here, we focus on another simulation package, ESPResSo (Extensible Simulation Package for Research on Soft Matter Systems), an open source software, which has been developed at the Institute for Computational Physics of the University of Stuttgart. ESPResSo is typically used for MD simulations of large scale coarse-grained (CG) models of soft-matter systems, and it includes a BBK-type discrete-time thermostat.

To demonstrate the performance of the G-JF thermostat for more complex soft systems, we simulated a bilayer membrane of  CG model lipids (see Fig.~5) using the ESPResSo package. Each lipid is modeled as a trimmer consisting of one hydrophilic and two hydrophobic beads of diameter $\sigma$, and the simulations are performed with no explicit solvent and with the Cooke-Kremer-Deserno force fields \cite{Cooke}. Specifically, all beads are subjected to a  short-range repulsive potential by applying a cut-off to a standard Lennard-Jones potential, which is vertically shifted such that
\begin{eqnarray}
V_{\rm rep}(r) & = & \left\{\begin{array}{ccc}
\displaystyle{4\varepsilon\left[\left(\frac{\sigma^\prime}{r}\right)^{12}-\left(\frac{\sigma^\prime}{r}\right)^{6}+\frac{1}{4}\right]} & , & r<r_c \\
0 & , & r\ge r_c\end{array}\right. \label{eq:repulsive}
\end{eqnarray}
where $\sigma^\prime=0.95\sigma$ for head-head and head-tail interactions, $\sigma^\prime=\sigma$ 
for tail-tail interactions, and $r_c=\sqrt[6]{2}\sigma^\prime$. The bonds connecting the intra-lipid beads are described by the FENE potential
\begin{eqnarray}
V_{\rm bond} & = & -\frac{1}{2}k_{\rm bond}r_\infty^2\ln\left[1-\left(\frac{r}{r_\infty}\right)^2\right] \, , \label{eq:V_bond}
\end{eqnarray}
with $k_{\rm bond}=30\varepsilon/\sigma^2$ and $r_\infty=1.5\sigma$. Each lipid is straightened by a harmonic spring potential between the head and the second tail bead, given by
\begin{eqnarray}
V_{\rm bend} & = & \frac{1}{2}k_{\rm bend}(r-4\sigma)^2 \, , \label{eq:V_bend}
\end{eqnarray}
where the bending stiffness $k_{\rm bend}=10\varepsilon/\sigma^2$. Finally, an attractive non-bonded interaction energy is introduced between any pair of hydrophobic tail beads. The attractive potential is given by
\begin{eqnarray}
V_{\rm attr} & = & \left\{\begin{array}{ccc}
-\varepsilon & , & r<r_c \\
-\varepsilon\cos^2\frac{\pi(r-r_c)}{2\omega_c} & , & r_c\le r < r_c+\omega_c \\
0 & , & r\ge r_c+\omega_c\end{array}\right.
\end{eqnarray}
We choose the parameters $\varepsilon=k_BT$ and $\omega_c=1.35\sigma$, which produce a bilayer membrane in the fluid state \cite{Cooke}.
Normalized friction coefficients, masses, and temperature are chosen to unity.
A snapshot of the bilayer membrane, taken from one of the simulations, is presented in Fig.~5.

\begin{figure}[t]
\begin{flushleft}
\scalebox{1.00}{\centering \includegraphics{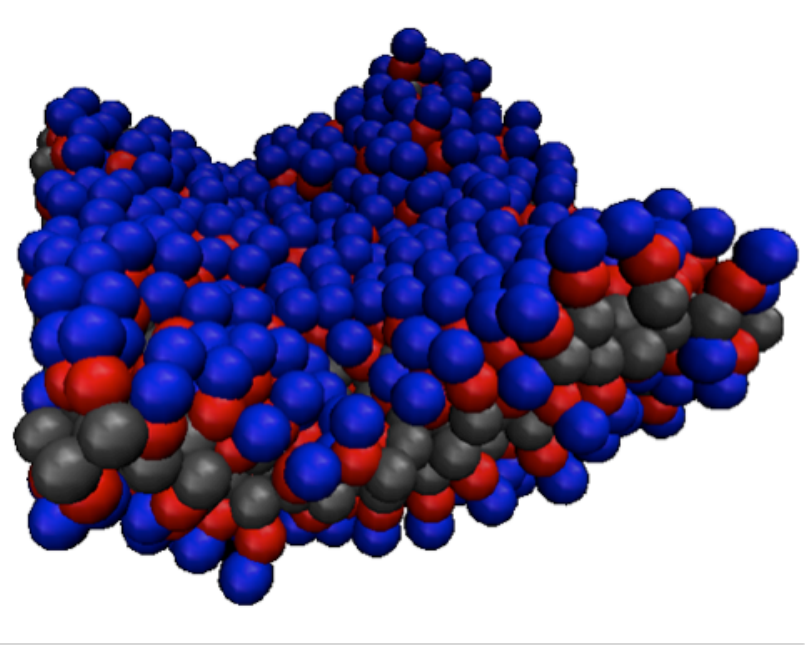}}
\end{flushleft}
\vspace{-0.5 cm}
\caption{An equilibrium snapshot consisting of 500 lipids. Head, first tail, second tail beads are, respectively, depicted in blue, red and grey.}
\label{fig:fig_5}
\end{figure}

\begin{figure}[t]
\begin{flushleft}
\scalebox{0.70}{\centering \includegraphics{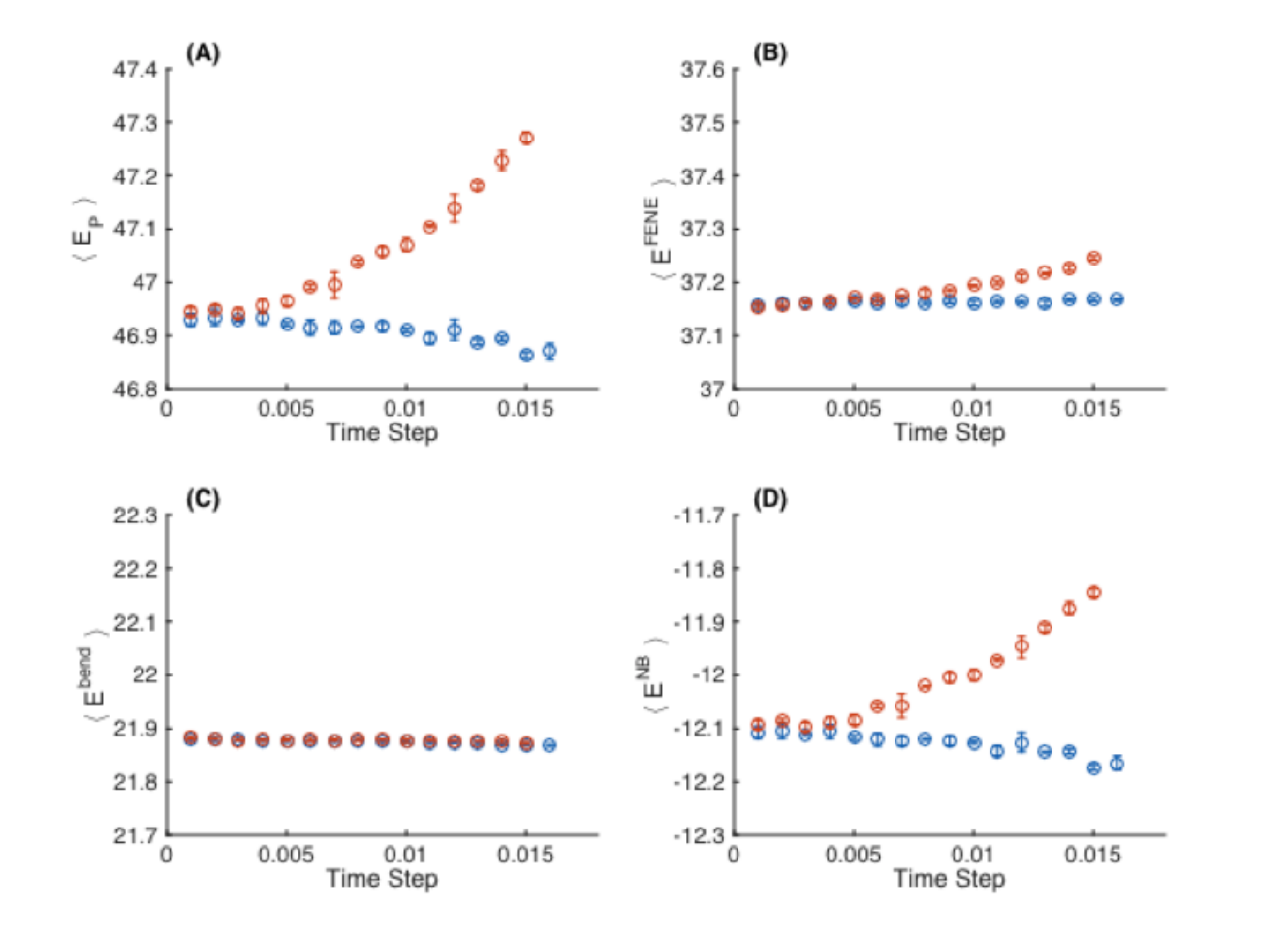}}
\end{flushleft}
\vspace{-0.5 cm}
\caption{The mean potential energy (A) and the separate contributions of the FENE bonds (B), bond-bending energy (C), and non-bonded interactions (D), as a function of the time step of the simulation. Red and blue symbols present, respectively, the results obtained with the Langevin thermostat currently installed in ESPREesSo, and those obtained when the thermostat is replaced by G-JF. Energeis are normalized per lipid.}
\label{fig:fig_6}
\end{figure}

\begin{figure}[t]
\begin{flushleft}
\scalebox{0.70}{\centering \includegraphics{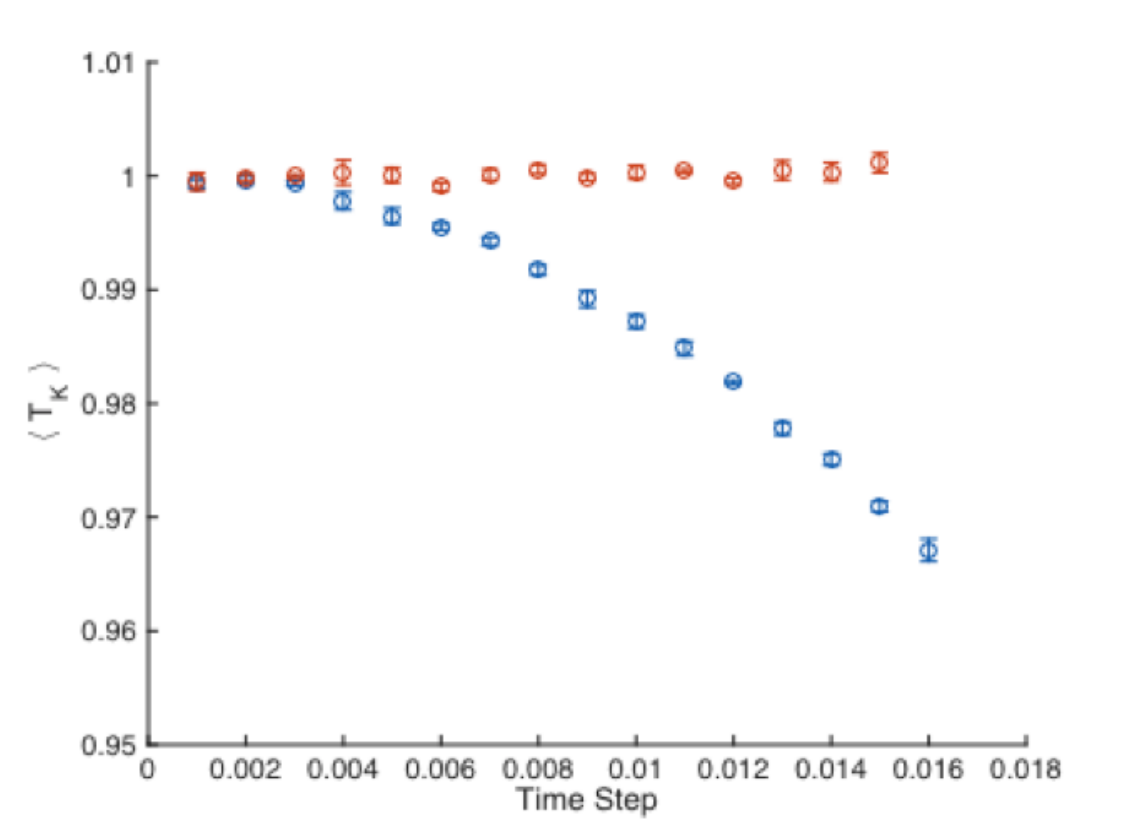}}
\end{flushleft}
\vspace{-0.5 cm}
\caption{The measured kinetic temperature as a function of $dt$. Red and blue symbols present, respectively, the results obtained with the Langevin thermostat currently installed in ESPREesSo, and those obtained when the thermostat is replaced by G-JF. Energies are normalized per lipid.}
\label{fig:fig_7}
\end{figure}

The bilayer membrane was simulated for 43,200 simulation time units, with integration time step $dt$ varying from $dt=0.001$ and up to the stability limit of the system ($dt\approx0.016$) in increments of $\Delta dt=0.001$. System-wide energy measurements were taken every 12 time units, and included the kinetic temperature $T_k=\frac{2}{9}\langle E_k\rangle/N$,
where $N$ is the number of lipids (each modeled with three beads),
total potential energy $\langle E_p\rangle$, and the separate contributions to $\langle E_p\rangle$ due to the FENE bonds (17), bond-bending energy (18) and non-bonded (NB) energy [sum of Eqs.~(16) and (19)].  The box size for the simulation was set to  $(17.6\sigma)^3$ (corresponding to nearly tensionless conditions), and was subjected to periodic boundary conditions. For each $dt$, two sets of simulations were performed: one with the Langevin thermostat currently used by ESPResSo, and the other with the G-JF integrator, which was implemented into the ESPResSo code. The results for $\langle E_p\rangle$ and its three constituting components are depicted in Fig.~6, as a function of the simulation time-step $dt$.  Shown error bars are here evaluated by the standard deviation of the results of three thirds of the total simulation time. Consistent with previous studies comparing the performance of various Langevin thermostats \cite{GJF2}, the results here also demonstrate that, unlike other methods, and over the entire stability range, the G-JF integrator does not create increasing artificial variations in the sampling statistics. We also reveal that with the thermostat currently implemented in ESPResSo, the error in the potential energy is caused mainly by the NB interactions, while a smaller error arises from the FENE bonds. The bond bending interactions seem to be accurately evaluated for all time steps. These observations can be understood considering the curvature of the interaction energies, which is largest for the repulsive pair potential (16), and smallest for the bond-bending interaction (18).

Figure 7 shows the computed results for the measure of kinetic temperature $T_k$, as a function of $dt$. The trends observed here are opposite to the ones shown in Fig.~6. The G-JF method leads to a decrease in $T_k$ with $dt$, while the ESPREesSo thermostat gives $k_BT_k/E_0=1$ for all simulated time steps. This feature has also been previously observed and discussed \cite{GJF1,GJF2,Eastwood,GJF3}. 
As mentioned in the introduction,
it stems from the fact that the discrete-time momentum $mv^n$ is not exactly conjugated to the coordinate $r^n$. Consequently, the kinetic temperature is not a good measure for high-quality statistical sampling -- see Ref.~\cite{GJF3}. In general, thermostats exhibiting correct $T_k$ in discrete time must produce errors in computed configurational thermodynamic quantities.

\section{Conclusion}
In conclusion, the G-JF Langevin thermostat has been tested on both simple linear and nonlinear oscillators, and it has been demonstrated that the expected exact statistics for linear systems is obtained. For nonlinear systems, we find some deviations for large time steps. These originate from the inherent time discretization of the deterministic force -- a feature common to all discrete-time numerical methods. The correct discrete-time implementation of the fluctuation relationship through the G-JF method is validated by the limit of large $\alpha$,
where the dynamics is dominated by noise and friction, and where G-JF gives near-perfect agreement with the continuous-time expectation. We have further implemented G-JF into the ESPREesSo molecular simulation package, and it has been applied for simulations of a CG implicit-solvent bilayer membrane. The simulation results presented here demonstrate, once again, that this newly developed integrator exhibits no shift in the values of measured configurational thermodynamic quantities with increasing simulation time steps. This allows one to run a simulation with considerably larger time steps, and provides the user with peace of mind about the accuracy of the configurational results.  The G-JF integrator is currently available within the LAMMPS simulation package, and we advise users of other popular suites, where older, considerably less accurate thermostats are implemented, to run simulations with caution and small time steps $dt$.

\section{Acknowledgments}
This work was supported in part by the Israel Science Foundation, Grant No.~1087/13, in part by the U.S.~Department of Energy, Grant No.~DE-NE0000536 000.

\end{document}